# 9.8-GW tabletop attosecond light source

**Lihui Meng,[1,‡] Xusheng Zhu,[1,‡] Lu Xu,[1,*] Lixin He,[1] Zan Nie,[1] Pengfei Lan,[1,†] and Peixiang Lu[1,‡]**

[1]*Wuhan National Laboratory for Optoelectronics and School of Physics, Huazhong University of Science and Technology, Wuhan 430074, China*

[*] *luxu_0909@hust.edu.cn*

[†]*pengfeilan@hust.edu.cn*

[‡]*lupeixiang@hust.edu.cn*

[‡]*these authors contributed equally to this work*

## Abstract

The generation of attosecond pulses (1 as=$10^{-18}$ s) has enabled real-time observation and manipulation of coherent electron dynamics, yet their low peak power has hindered the development of advanced attosecond pump-probe spectroscopy and attosecond nonlinear metrology. Here we overcome this limitation by generating 2.86-μJ, 292-as isolated attosecond pulses (IAPs) with a peak power of 9.8 GW—the highest pulse energy and peak power reported for a tabletop isolated attosecond source. This is achieved by combining a 2.4 TW, few-cycle (8.3 fs) two-color synthesizer with a loose focusing geometry that enables macroscopic phase-matching. The synthesizer features a stabilized carrier-envelope phase and an actively synchronized relative time delay between the two-color channels, ensuring high stability and reproducibility. This high-power tabletop attosecond source will open up nonlinear effect experiments that were previously inaccessible with lower-power IAPs, establishing a foundation for advanced attosecond spectroscopy and nonlinear metrology.

## Introduction

Since the first demonstration of attosecond pulses via high-order harmonic generation (HHG) [1,2], attosecond science has profoundly impacted the study of ultrafast phenomena. Time-resolved attosecond spectroscopy has enabled the tracking of ultrafast electron dynamics inside atoms, molecules, and more recently in solids and liquids [3-6], greatly advancing our understanding of the transient dynamics of laser–matter interactions. Extensive efforts have focused on temporally compressing the attosecond pulse, leading to various gating methods, including amplitude gating driven by few-cycle pulses [1,7], polarization gating [8], double optical gating [9], and two-color gating [10-12]. Over the past decades, attosecond pulse durations have been progressively compressed from several hundred attoseconds to below 50 attoseconds [13-17].

Despite these impressive advancements in pulse compression, the application of attosecond pulses remains constrained by the low pulse energy and peak power of the current attosecond sources [18]. In particular, attosecond nonlinear effects, e.g., multi-photon absorption or ionization, second harmonic generation, require attosecond pulses with sufficiently high peak intensity to facilitate nonlinear processes on their intrinsic timescale, whereas attosecond-pump-attosecond-probe spectroscopy demands an adequate photon flux of attosecond pulses to capture correlated electron dynamics with attosecond temporal resolution [19]. The peak power of X-ray attosecond pulses generated by free-electron lasers (FEL) has exceeded the terawatt scale [20]. Nevertheless, the kilometer-long facility size and substantial costs pose significant challenges to widespread and flexible use as a readily accessible tool for cutting-edge fundamental research. Therefore, developing a tabletop, high-power attosecond source remains a critical and urgent requirement in ultrafast science [21].

Parametric waveform synthesis has been demonstrated to control the HHG and produce tabletop isolated attosecond pulses (IAPs) with customized optical fields [22]. Combining two-color field synthesis with energy-scaling techniques has previously achieved 2.6-gigawatt (GW) IAPs with a duration of 500 as [23]. This approach was later extended to multichannel optical synthesizers, which successfully generated 226-as IAPs, albeit with a reduced peak power of 1.1 GW [24,25]. However, those implementations either suffered from residual timing jitters or carrier-envelope phase (CEP) fluctuations, or faced considerable difficulties in handling the complex multichannel synthesizer, limiting their stability and reproducibility. Producing high-power IAPs faces several principal challenges. First, it requires temporally confining HHG to a sub-cycle time scale of the driving field. Second, owing to the low conversion efficiency of HHG ($10^{-4} \sim 10^{-5}$), the pulse energy of the driving laser needs to reach tens of millijoules, with an ultrashort duration and customized optical field. Third, improving the IAP yield requires the coherent buildup of the HHG, i.e., macroscopic phase-matching [26]. Simultaneously overcoming these difficulties remains a formidable hurdle for current state-of-the-art technology.

In this work, we surmount these challenges by employing an experimental architecture for efficient IAP generation with a terawatt few-cycle two-color synthesizer. A cascaded post-compression was implemented to compress the fundamental pulse from 25 fs to 8.3 fs with a maximum pulse energy of 20 mJ, which was synthesized with a frequency-doubled auxiliary pulse. To achieve precise sub-cycle control over HHG, we meticulously controlled the waveform of the two-color synthesizer by stabilizing the CEP of the few-cycle driving laser and actively synchronizing the relative timing delay within the two-color synthesizer using a balanced optical cross-correlator (BOC). Meanwhile, by carefully designing the phase-matching of HHG with a loosely focusing geometry, we generated a supercontinuum HHG spectrum, resulting in IAPs with pulse energies up to 1.64 μJ in Krypton (Kr), 2.86 μJ in Xenon (Xe) and 0.72 μJ in Argon (Ar). The pulse durations were measured to be 263 as, 292 as and 230 as, respectively. Delivering a maximum peak power of 9.8 GW, this tabletop attosecond source sets a new benchmark, paving the way for routine access to attosecond nonlinear optics and pump-probe spectroscopy.

# Results

## Terawatt few-cycle two-color synthesizer

The detailed experimental setup is illustrated in Fig. 1a. It consisted of a 50-Hz front-end Ti:sapphire laser (45 mJ, 25 fs, 800 nm), whose output was partially split into two parts (36 mJ and 9 mJ, respectively). A cascaded post-compressor, housed in a vacuum chamber, was implemented to compress the 36 mJ Ti:sapphire laser pulses while avoiding deleterious ionization and severe degradation of the beam profile. The output from the cascaded post-compressor was an 8.3-fs few-cycle pulse (Fig. 1b) with a maximum pulse energy of 20 mJ, resulting in a peak power of 2.4 TW (see the "Cascaded post-compression" section in Methods), which was employed as fundamental laser pulses for the two-color synthesizer. Simultaneously, the 9-mJ laser pulses from the front-end were frequency-doubled in a 300-μm-thick β-barium borate (BBO) crystal, producing 2-mJ, 410-nm auxiliary laser pulses for the two-color synthesizer. Owing to the bandwidth limitation of the frequency-doubling process, the resulting pulse duration was 40 fs. Notably, the residual Ti:sapphire laser pulse after the frequency-doubling process, once separated by a dichroic mirror, served as the perturbation laser for the all-optical measurement of the attosecond pulse duration (see the "All-optical characterization of IAPs" section in Methods).

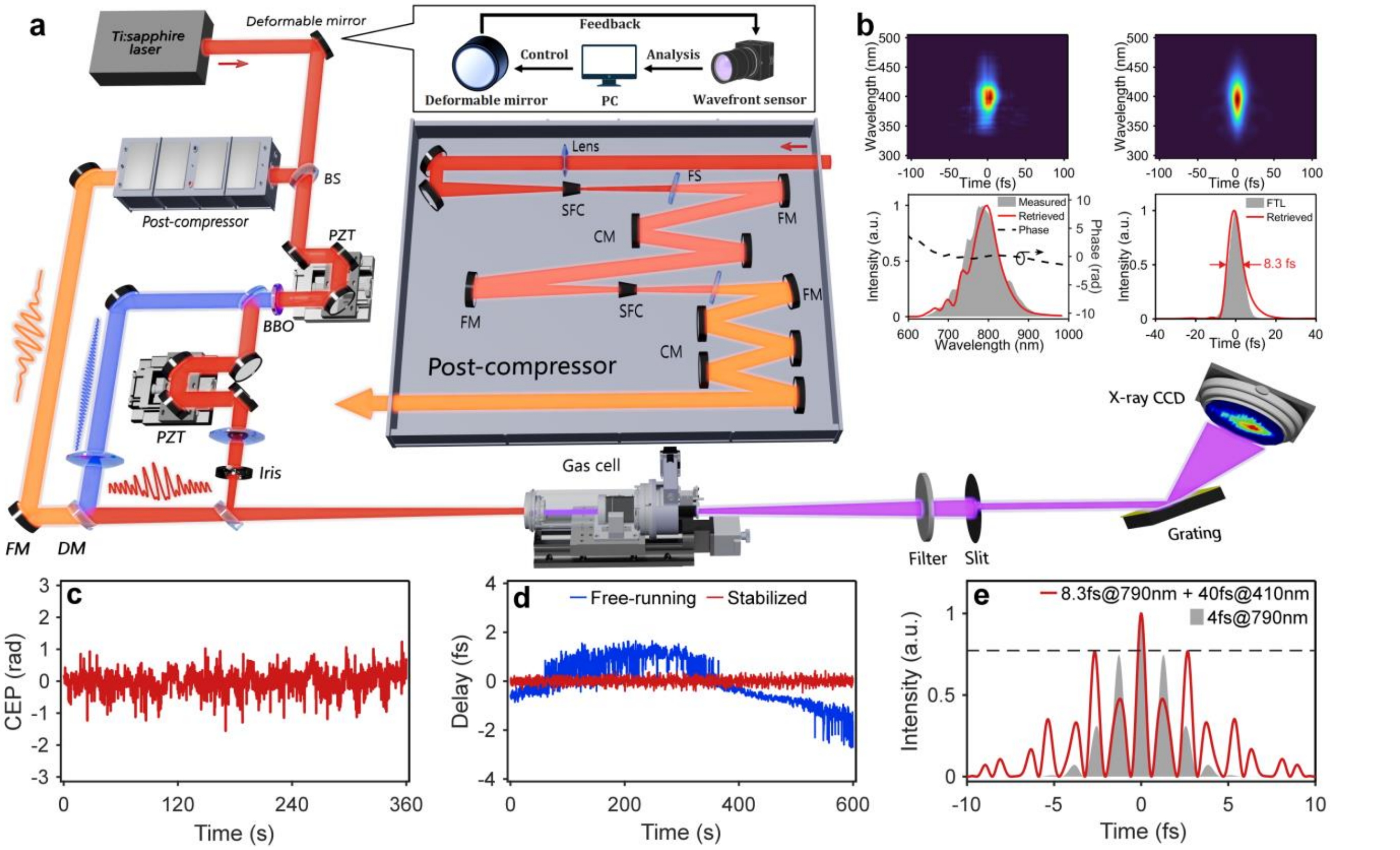

**Fig. 1. Experimental setup for the generation of GW-class IAPs. a**, Terawatt few-cycle two-color synthesizer. BBO, β-barium borate; PZT, piezo-transducer-actuated stage; CM, chirped mirror; DM, dichroic mirror; FM, focusing mirror. **b**, SHG-FROG characterization of the output pulse from the cascaded post-compressor. Measured (top left) and retrieved (top right) SHG-FROG traces. Bottom left: Retrieved spectrum (red solid line) and spectral phase (black dash line), measured spectrum (gray fill). Bottom right: Retrieved temporal profile (red solid line), which indicated a pulse duration of 8.3 fs (full width at half-maximum). The Fourier-transform limit pulse duration is 7.8 fs (FWHM, gray fill). **c**, CEP shift of the 50-Hz front-end Ti:sapphire laser, with an evaluated root mean square (RMS) error of 318 mrad. **d**, Delay jitter of the two-color synthesizer. RMS delay jitter was improved from 888 as (free-running) to 116 as after stabilization with a balanced optical cross-correlator (BOC). **e**, The intensity profiles of the two-color synthesizer (red solid line). The intensity ratio between the main and second peaks resembled that of the 4-fs, 790-nm one-color field (gray fill).

The electric field of the two-color synthesizer can be expressed as:

$$\mathrm{E(t)} = \mathrm{E}_1 \exp\left(-2ln2\frac{t^2}{\tau_1^2}\right)\cos(\omega_1 t + \phi_1) + E_2 \exp\left(-2ln2\frac{(t-\delta t)^2}{\tau_2^2}\right)\cos(\omega_2(t-\delta t) + \phi_2) \quad (1)$$

where $\mathrm{E_i}$, $\tau_i$, $\phi_i$ (i=1, 2 for the fundamental and auxiliary pulses, respectively) are the amplitudes, pulse durations and CEPs, and $\delta t$ denotes the relative time-delay between the two-color channels. The fundamental and auxiliary pulses both originate from the same front-end Ti:sapphire laser, and the frequency-doubling process introduces no additional CEP jitter [27]. Consequently, the optical field of the synthesizer can be fully stabilized and precisely tailored by controlling the CEP $(\phi_1)$ and relative time-delay $(\delta t)$. As illustrated in Fig. 1c and Fig. 1d, the developed two-color synthesizer not only exhibited CEP stabilization but also possessed the capability to manipulate relative time delays. By tailoring the electric field, the synthesized two-color field becomes equivalent to a 4-fs single-color laser pulse (Fig. 1e), which enables gating the HHG process within a sub-cycle timescale and produces IAPs.

## Isolated attosecond pulse generation

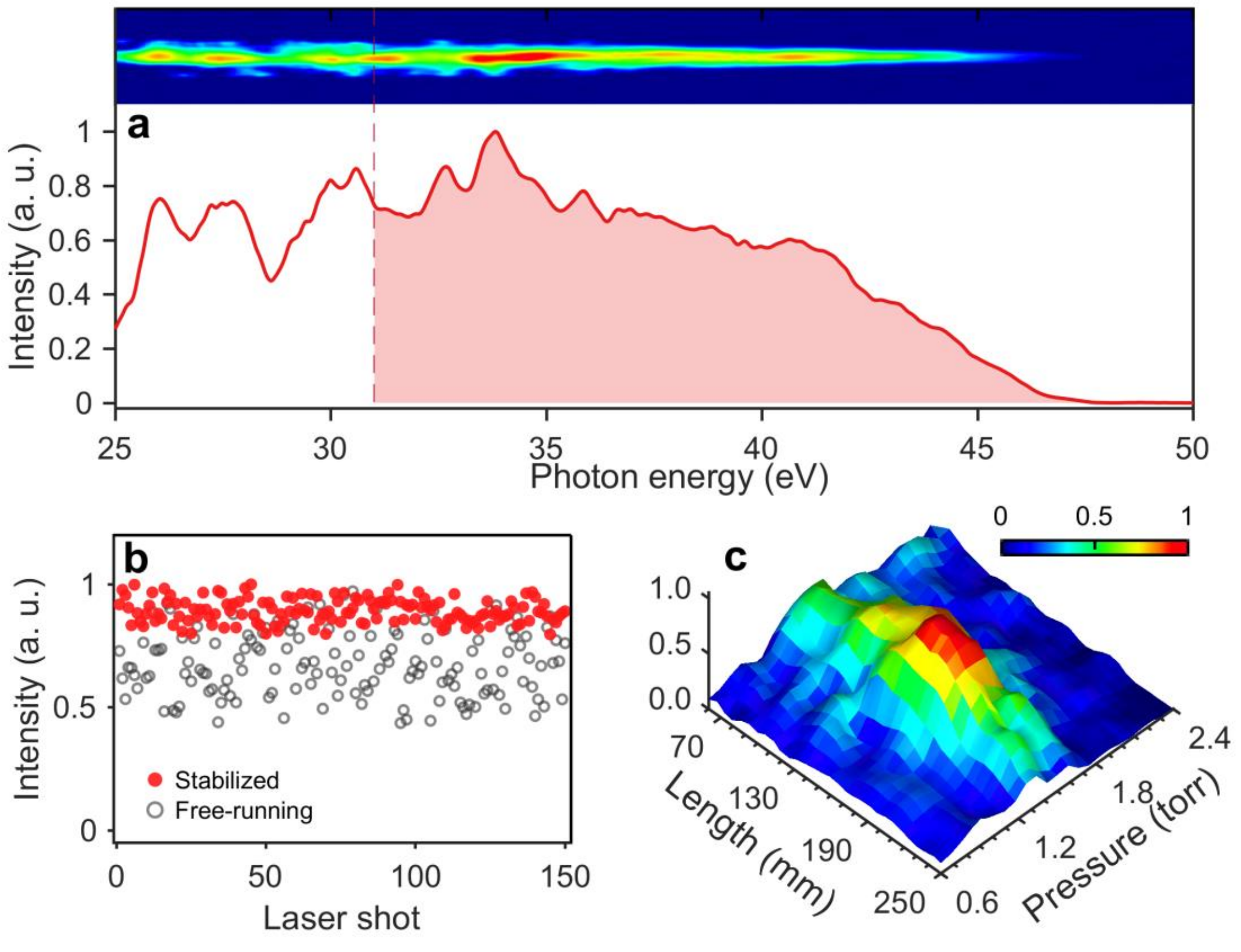

**Fig. 2. Isolated attosecond pulse generation in Krypton. a,** Single-shot high harmonic spectrum, experimentally acquired under the excitation of terawatt few-cycle two-color synthesizer. It clearly revealed a supercontinuum in the cutoff region from 31 to 48 eV. **b,** The measured fluctuations of the continuum high harmonics yield with and without the stabilization of the synthesizer. **c,** The integrated yield of the continuum high harmonics under varying gas pressure and interaction length.

According to the energy scaling law of HHG [28,29], the pulse energy of the driving laser can be fully utilized to increase the HHG yield by enlarging the focal spot size while keeping its peak intensity below the saturation level of the interacting gas. It meanwhile results in larger interaction volumes. To maintain the phase matching under expanded interaction volumes, the medium length must be proportionally increased to compensate for the reduced geometric Gouy phase shift.

In the experiments, the synthesizer laser beam was focused on the Kr gas cell with a 9-m focusing geometry. Under the driving of solely fundamental laser pulse (18 mJ, 8.3 fs), the spectrum exhibited discrete odd-order harmonics. When the terawatt few-cycle two-color synthesizer was applied, where the intensity ratio between the auxiliary pulse and the fundamental pulse was 3%, a distinct continuous spectral range emerged (Fig. 2a). Meanwhile, by scanning both the pressure of Kr gas and the interaction length, we optimized the macroscopic phase-matching conditions for HHG to achieve high yield and a broadband continuum spectrum (Fig. 2c). Following this optimization, a continuous HHG spectrum spanning from 31 to 48 eV, as depicted in Fig. 2a, was obtained at a Kr gas pressure of 1.34 torr and an interaction length of 165 mm. From the observed HHG spectrum (Fig. 2a), the effective interaction intensity was estimated to be $1.1 \times 10^{14}$ (W/cm$^2$) based on the cutoff law [30], corresponding to an ionization probability of less than 4% from the Ammosov-Delone-Krainov model [31].

Consistent generation of high harmonics critically relied on the active stabilization of the synthesizer. To experimentally evaluate the stability of the HHG output, we recorded 150 consecutive single-shot spectra. As illustrated in Fig. 2b, in the absence of active CEP and delay stabilization, the integrated continuous high harmonics yield displayed pronounced fluctuations ranging from 0.4 to 1.0 times the maximum intensity. By implementing both CEP and delay stabilization, the yield fluctuation significantly diminished to 4.9% RMS, demonstrating a highly stable tabletop attosecond source suitable for long-term measurements.

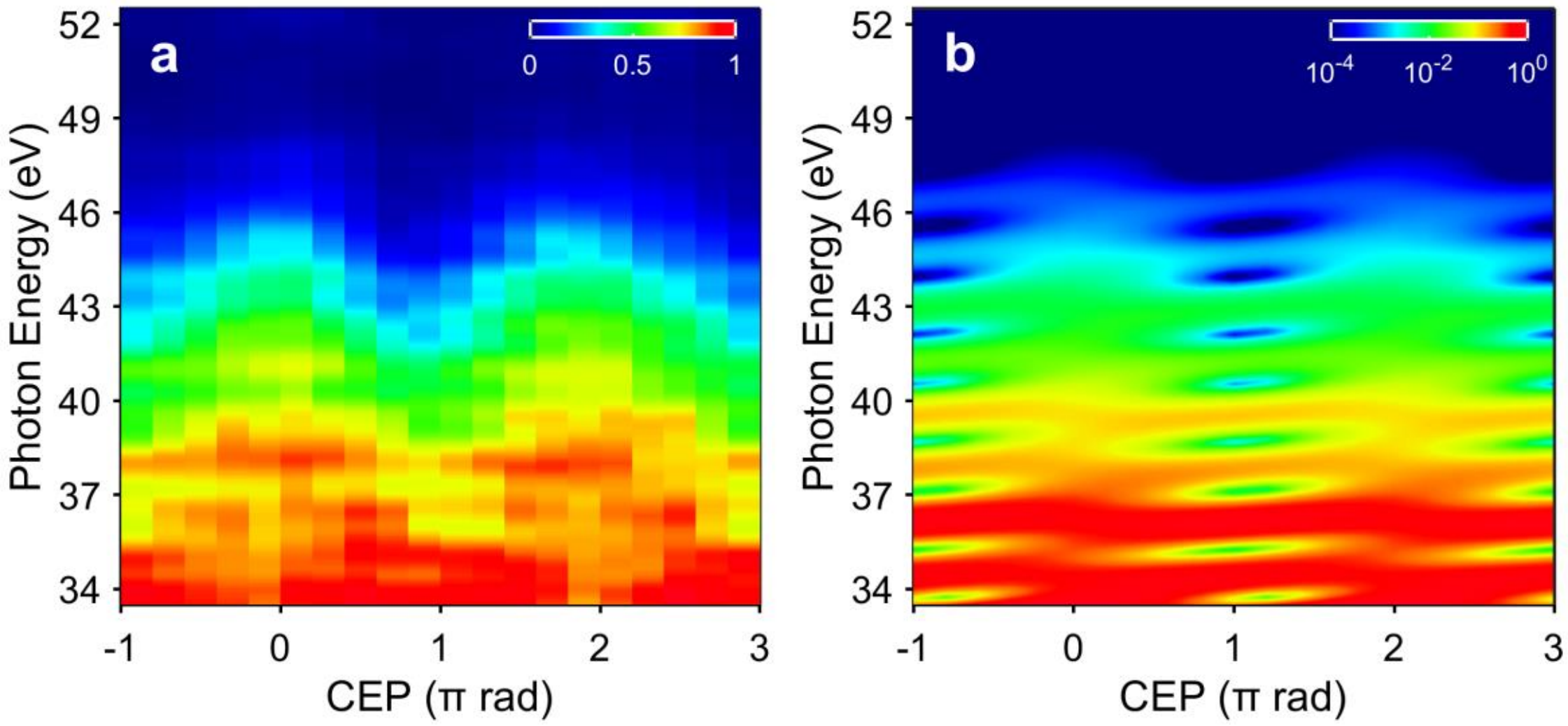

**Fig. 3. CEP dependence of the measurements and simulations of high harmonics spectrum generated in Kr. a**, Experimentally measured HHG spectrum by scanning the relative CEP in steps of 0.2π rad. **b**, Simulated HHG spectrum as a function of the relative CEP, reproducing the key features observed in the experiment.

The CEP dependence of HHG was demonstrated by recording high harmonic spectra by scanning the CEP in steps of 0.2π rad. In Fig. 3a, similar intensity distributions of the HHG spectrum repeated every 2π rad of CEP shift. Furthermore, when the CEP was zero, the harmonic spectra exhibited the highest cutoff photon energy and the broadest continuum. This characteristic was corroborated by consistent results from experiments (Fig. 3a) and simulations (Fig. 3b). It was evident that complete command over the synthesized waveforms facilitated precise modulation of electron dynamics and the reliable generation of IAPs.

## Temporal and energetic characterization of generated IAPs

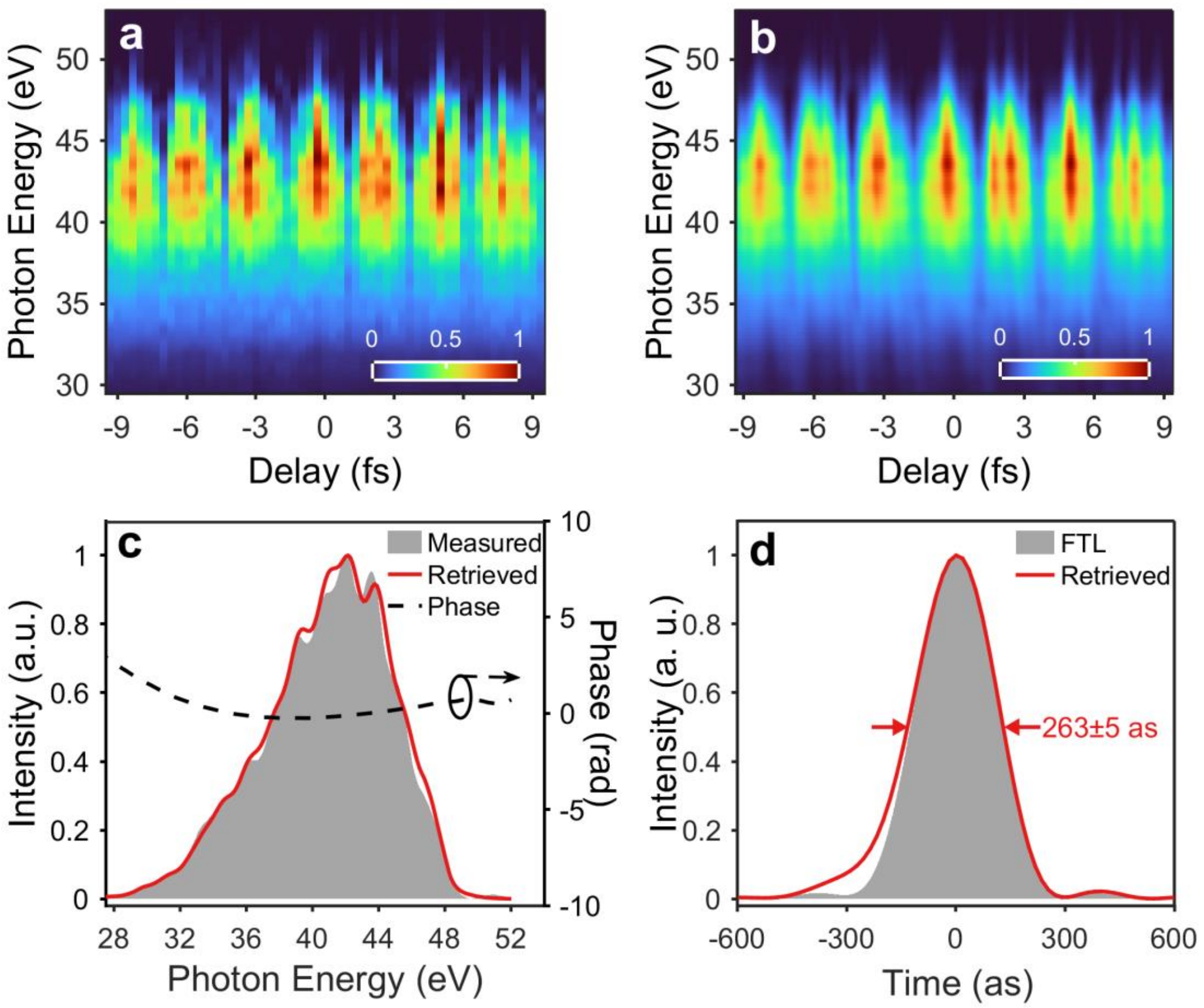

**Fig. 4. Temporal characterization of the generated IAPs in krypton by all-optical FROG. a,** Measured two-dimensional all-optical FROG trace. **b,** Retrieved all-optical FROG trace. **c**, Reconstructed spectrum (red solid line), reconstructed spectral phase (black dash line). The gray filled spectrum shows the continuous high harmonics obtained by filtering the experimentally measured spectrum in Fig. 2a using a Parylene-N filter, agreeing well with the retrieved results. **d**, Retrieved temporal profile (red solid line) and pulse profile in the Fourier-transform limit (gray fill).

We adopted the all-optical Frequency-Resolved Optical Gating (FROG) method [32], which has been demonstrated to yield consistent results with the attosecond streaking camera [32,33,25], to characterize the temporal profile of the IAPs. As illustrated in Fig. 1a, a weak field from the residual pump of the frequency-doubling process in the auxiliary pulse acted as a gate pulse to perturb the HHG process. The perturbed high harmonic spectra, as a function of the relative time delay between the two-color synthesizer and the weak gate pulse, represent a phase modulation of the HHG (i.e., the all-optical FROG trace). We employed a Parylene-N filter to process the measured spectra prior to the subsequent reconstruction of the IAP duration. Since the Parylene-N filter exhibited significant opacity at wavelengths below 30 eV, the continuous spectrum obtained via this filtering approach closely aligned with the experimental results (Fig. 2a). Fig. 4a demonstrates the measured all-optical FROG trace recorded at relative time delays in increments of 334 as. It reveals a delay-dependent modulation characterized by a period of approximately 2.7 fs, consistent with the optical cycle of the gate pulse. The solid red line and the dashed black line in Fig. 4c depict the retrieved spectrum and the spectral phase, respectively. The corresponding temporal pulse profile for the retrieved pulses is shown in Fig. 4d (see the "All-optical characterization of IAPs" section in Methods). The pulse duration was 263 as (full width at half-maximum, FWHM).

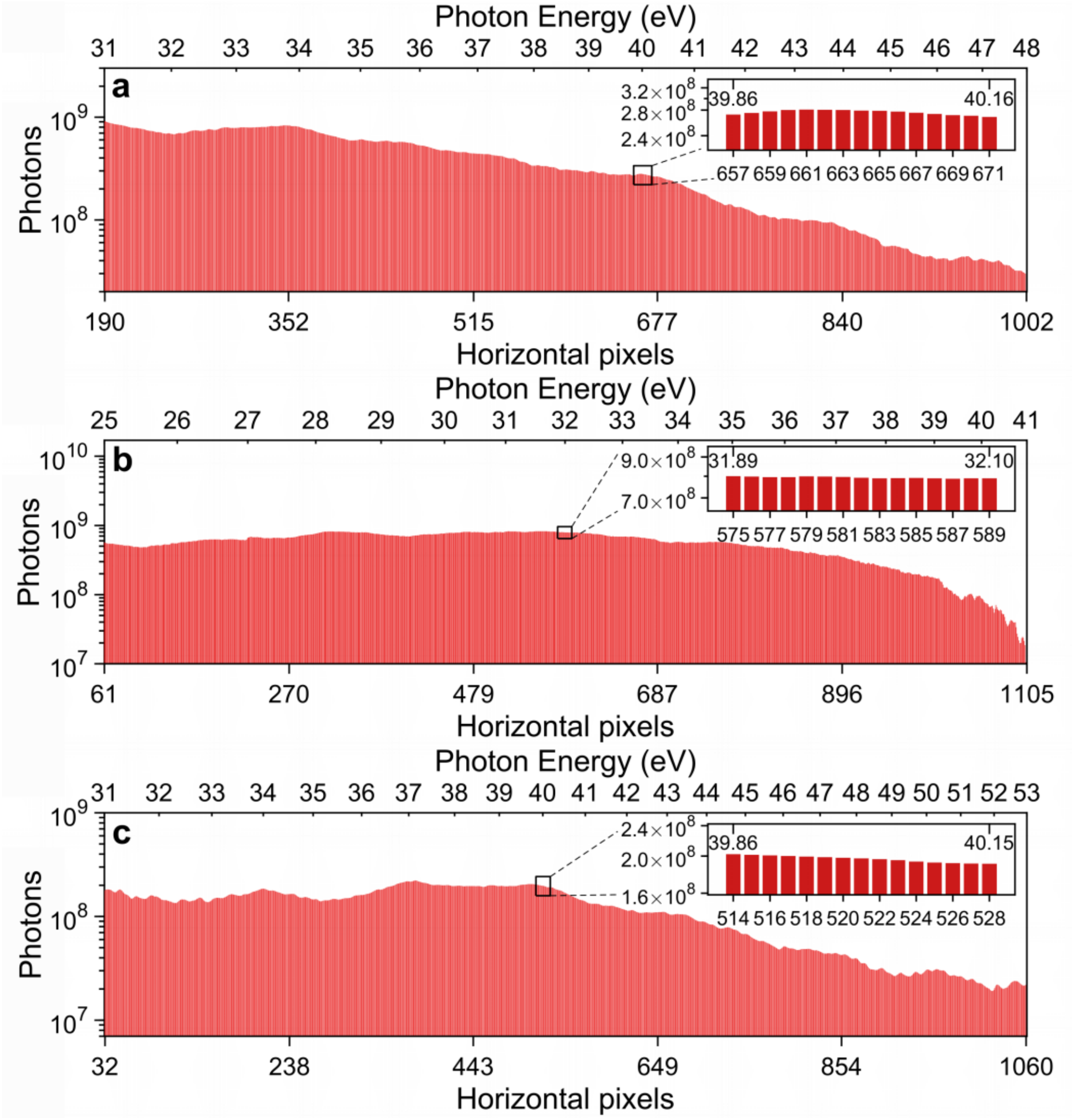

**Fig. 5. The generated photon counts in HHG.** **a**, Photon counts within a single-shot harmonic continuum (ranging from 31 to 48 eV) generated through HHG of Kr gas. **b**, Photon counts within a single-shot harmonic continuum (ranging from 25 to 41 eV) generated through HHG of Xe gas. **c**, Photon counts within a single-shot harmonic continuum (ranging from 31 to 53 eV) generated through HHG of Ar gas.

To calibrate the absolute energy of the generated IAPs, we employed an X-ray charge-coupled device (CCD) detector (PIXIS-XO 400B, Princeton Instruments) for single-shot photoelectron counting [34,35]. The total X-ray CCD counts are then converted to the absolute photon count per pulse ($N_{ph}$) using the following equation:

$$N_{ph} = \frac{S_{CCD} \cdot \sigma}{\eta_{QE} \cdot (E_{ph}/3.65\ \mathrm{eV}) \cdot \eta_g \cdot T_f \cdot T_s} \quad (2)$$

Here, $S_{CCD}$ denotes the cumulative readout counts of the detector, σ is the number of electrons corresponding to a single digital count unit; thus, $S_{CCD} \cdot \sigma$ yields the aggregate number of electrons collected by the detector. $\eta_{QE}$ represents the wavelength-dependent quantum efficiency of the detector, $E_{ph}$ is the photon energy inherent in the attosecond pulse, and the ratio $E_{ph}$/3.65eV is the number of electrons liberated per incident photon, governed by the 3.65 eV bandgap energy of the silicon substrate. Leveraging these parameters, we computed the photon counts, in accordance with the number of photons received by each pixel of the detector (Fig. 5). By summing the photon counts of a single-shot high harmonic continuum spanning from 31 to 48 eV (see the "Energy characterization of IAPs" section in Methods), the absolute energy of the single-shot IAPs was ascertained to be 1.64 μJ. Ultimately, considering the pulse duration of 263 as, it attained a peak power of 6.2 GW.

## Generation and characterization of IAPs in Xenon and Argon

To produce IAPs with higher pulse energy, we replaced the interaction medium with Xe gas, which has a lower ionization potential, and accordingly expended the focal length to 12 m to take advantage of the maximum pulse energy of fundamental laser (20 mJ) for driving HHG. After optimizing phase-matching conditions, we achieved an HHG spectrum with the continuum from 25 to 41 eV [see Fig. 6 (a)] at a lower gas pressure (approximately 0.6 torr) and a longer interaction length (around 206 mm) compared to those used in Kr gas. Because of the smaller ionization energy of Xe, the cutoff photon energy was smaller than that generated in krypton.

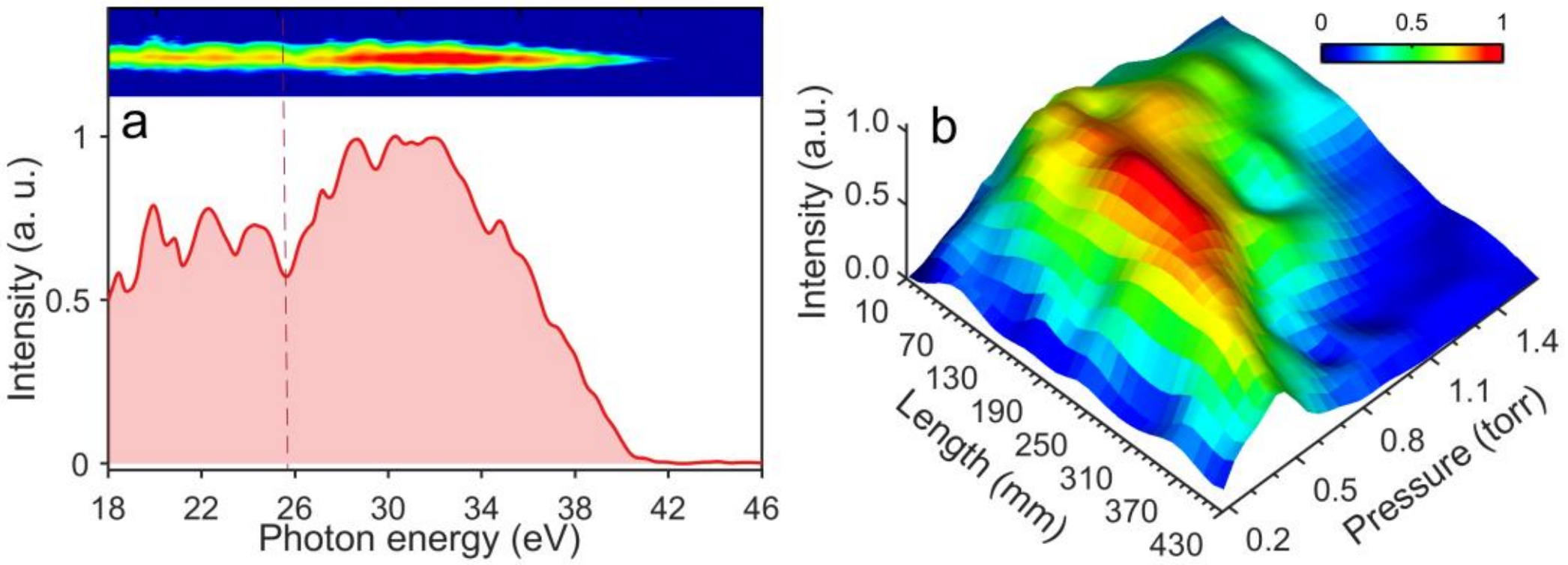

**Fig. 6. Isolated attosecond pulse generation in Xenon. a**, Single-shot high harmonic spectrum, experimentally acquired under the excitation of terawatt few-cycle two-color synthesizer. It clearly revealed a supercontinuum in the cutoff region from 25 to 41 eV. **b**, The integrated yield of the continuum high harmonics under varying gas pressure and interaction length.

A beryllium filter was employed to extract a continuous spectrum from the trace obtained through all-optical FROG measurements. As shown in Fig. 7, the pulse duration retrieved from the FROG trace is 292 as (FWHM). Using the same calibration methodology for attosecond pulse energy as in Kr, the pulse energy of IAPs in Xe was determined to be 2.86 μJ (Fig. 5b), and thereby the peak power reached 9.8 GW.

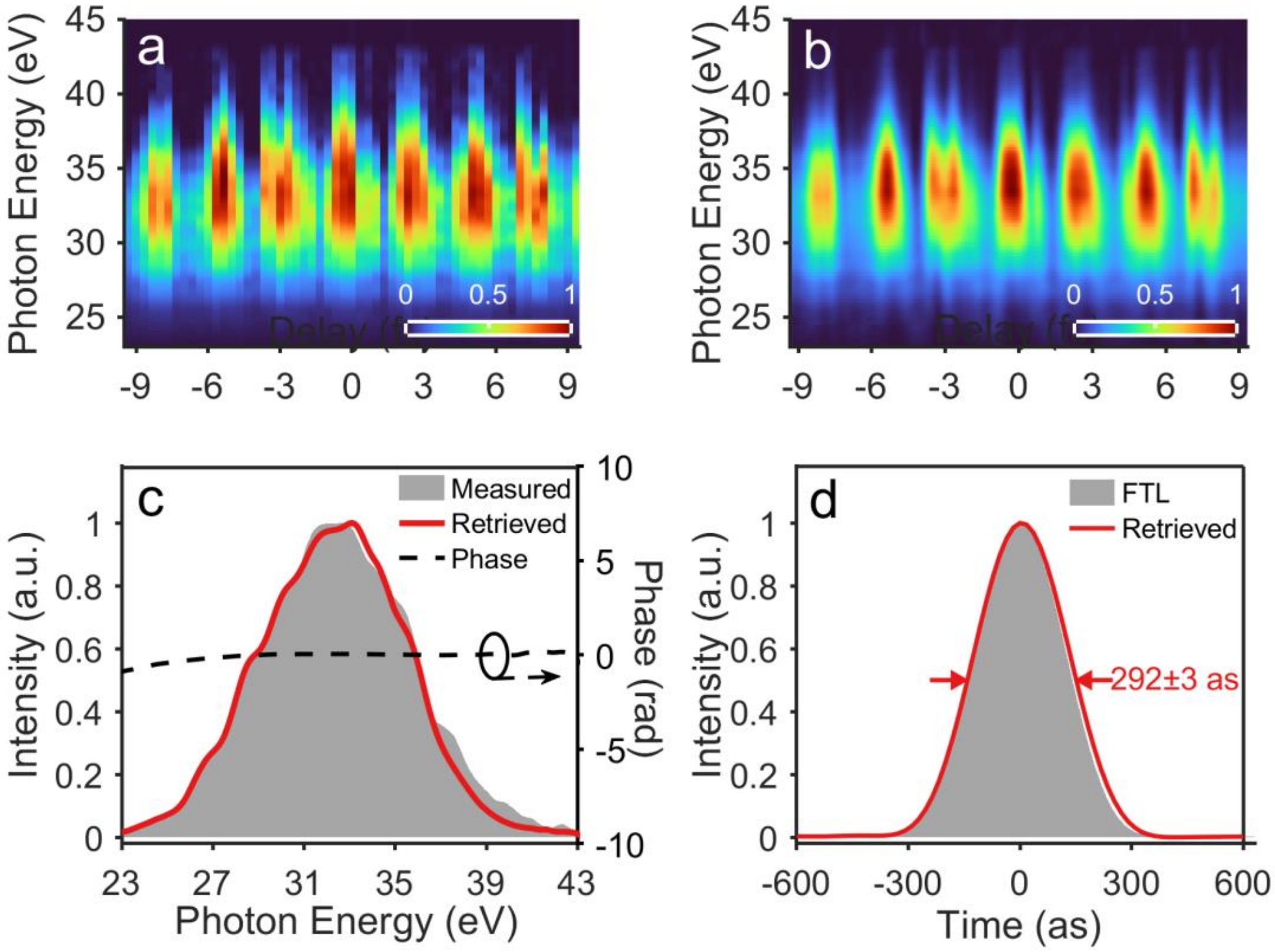

**Fig. 7. Temporal characterization of the generated IAPs in Xenon by all-optical FROG. a**, Measured two-dimensional all-optical FROG trace. **b**, Retrieved all-optical FROG trace. **c**, Reconstructed spectrum (red solid line), reconstructed spectral phase (black dash line). The gray filled spectrum shows the continuous high harmonics obtained by filtering the experimentally measured spectrum in Fig. 6a using beryllium filter. **d**, Retrieved temporal profile (red solid line) and the pulse profile in the Fourier-transform limit (gray fill).

In addition, to generate IAPs with shorter pulse duration, we substituted the interaction medium with Ar gas, which has a higher ionization potential, and accordingly reduced the focal length to 7.5 m for driving the HHG. After optimizing phase-matching conditions, we achieved an HHG spectrum with the continuum range from 31 to 53 eV, at a relatively higher gas pressure (approximately 2.1 torr) and a shorter interaction length (around 155 mm) compared to those used in Kr gas, as shown in Fig. 8a.

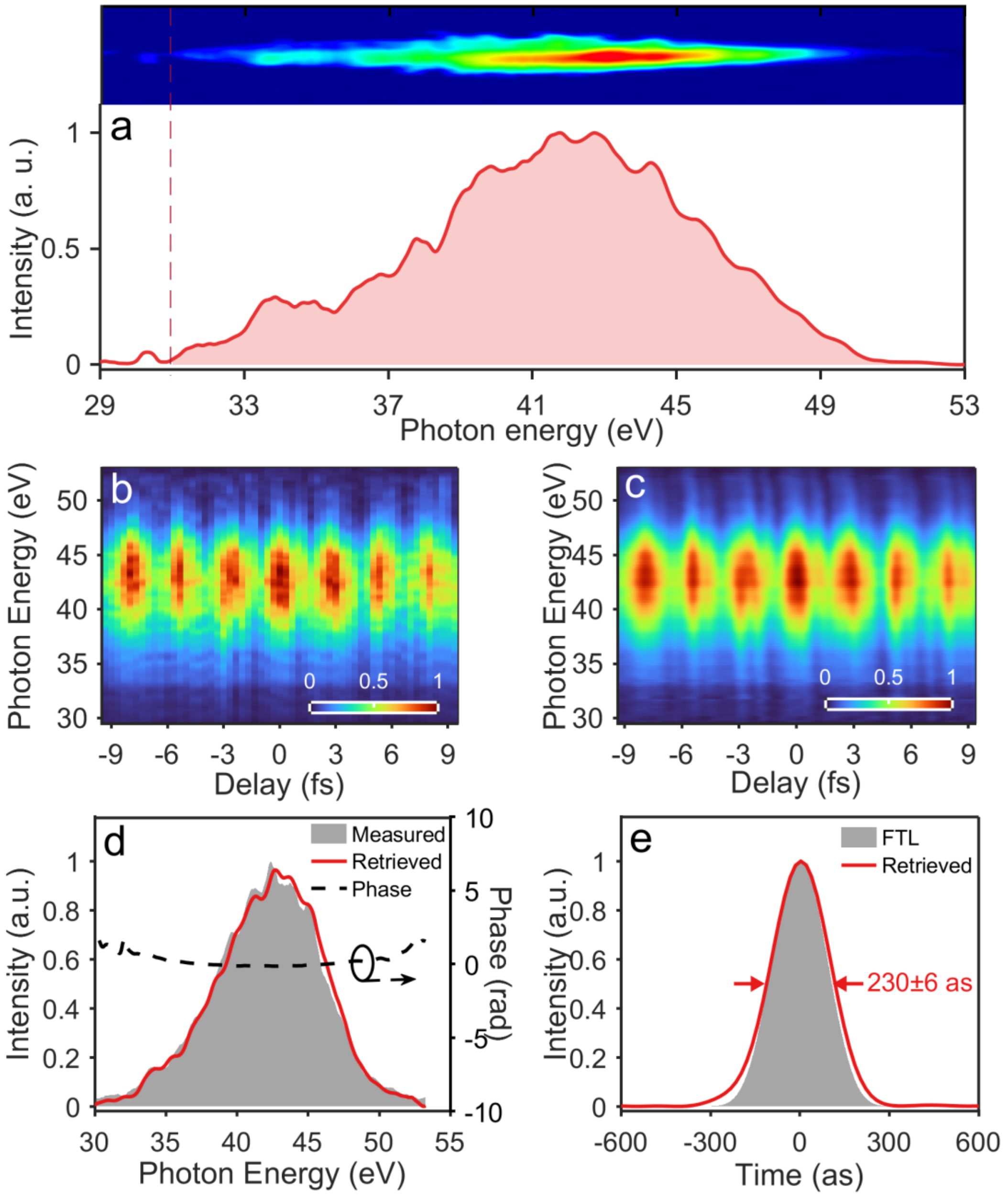

**Fig. 8. Experimental results of IAPs in argon gas. a**, Single-shot high harmonic spectrum, experimentally acquired under the excitation of terawatt few-cycle two-color synthesizer with a 7.5-m loosely focusing geometry. **b**, Measured two-dimensional all-optical FROG trace. **c**, Retrieved all-optical FROG trace. **d**, Reconstructed spectrum (red solid line), reconstructed spectral phase (black dash line). The gray filled spectrum shows the continuous high harmonics obtained by filtering the experimentally measured spectrum in Fig. 8a using Parylene-N filter. **e**, Retrieved temporal profile (red solid line) and the pulse profile in the Fourier-transform limit (gray fill).

Notably, the macroscopic harmonic yield within the high-energy range (> 40 eV) was larger than that in the low-energy plateau (25–40 eV), owing to the optimized generation efficiency of the cutoff harmonics. A Parylene-N filter was employed to extract a continuous spectrum from the trace obtained through all-optical FROG experimental measurements, yielding a retrieved pulse duration of 230 as (FWHM). Using the same calibration methodology for attosecond pulse energy as in Kr (Fig. 5c), the pulse energy of IAPs in Ar was determined to be 0.72 μJ, and thereby the peak power reached 3.1 GW.

## Discussion

It has already been shown that the HHG yield is dictated by a wavelength scaling law [36,37], which leads to a substantial decrease of HHG if a driving pulse of longer wavelength is used. However, in this work, by constructing the terawatt few-cycle two-color synthesizer in the near-infrared (790 nm) and visible regimes (410 nm), our strategy circumvented the wavelength-induced penalty and achieved a HHG generation efficiency of $1.3 \times 10^{-4}$. Moreover, the auxiliary pulse was obtained by the frequency-doubling process with a nonlinear crystal in our strategy, which significantly simplifies the synthesizer and enhances the robustness of the optical system.

High-power IAPs provide more feasible and convenient experimental conditions for several important applications. First, the multi-GW IAPs can induce nonlinear attosecond optics effects, e.g., two-photon ionization of atoms and molecules or second harmonic generation, and realize probing the ultrafast dynamics with full attosecond pump-probe spectroscopy. Second, current research on high-speed imaging based on FEL pulses suggests that it is capable of imaging the nanoscale structure in a single-shot exposure if the photon count reaches at least $10^{11}$–$10^{12}$ [38,39]. Remarkably, the microjoule-level IAPs fall within this range. Because of the ultrashort duration, the single-shot image with IAPs can capture the minute structural variations at the nanoscale. Nevertheless, achieving attosecond temporal resolution while preserving nanoscale spatial resolution will ultimately rely on the development of advanced broadband ultrafast imaging algorithms and techniques [40,41].

In conclusion, a stable experimental platform capable of generating IAPs at the 10-GW level has been demonstrated. To the best of our knowledge, both the pulse energy and peak power represent the highest values reported thus far for the laser-driven tabletop attosecond sources. The 10-GW level attosecond laser source will offer a breakthrough for research in nonlinear attosecond optics, attosecond pump-probe spectroscopy, and high-speed imaging.

## Materials and methods

### Cascaded post-compression.

The entire cascaded post-compressor was enclosed within a vacuum chamber (maintained at $10^{-4}$ mbar) to compress the 36 mJ Ti:sapphire laser pulses from the front-end to a few-cycle pulse duration while avoiding deleterious ionization and severe beam-profile degradation. We focused the input laser pulses with a 4-m focal length into the first post-compression stage. A spatial filter cone (SFC) placed near the focal spot mitigated spatial intensity modulations to eliminate self-focusing that occurred in the subsequent fused silica plate. The SFC had entrance and exit diameters of 1 mm and 0.3 mm, respectively, and a 2-mm-thick fused silica plate was placed 1.2 meters behind the focus to broaden the spectrum to 720-870 nm. The peak intensity of the input laser pulse was around 1.4 TW/cm$^2$ and the B-integral induced in the plate was calculated as 5.1. After collimation with a concave mirror of a 3-m focal length, the spectrally broadened laser pulse carried positive dispersion, which was the cumulative residual dispersion introduced by the fused silica plate, self-phase modulation, and optical components; this dispersion was compensated by a pair of CMs with a total chirp of -100fs$^2$. The output from the first post-compression stage was spectrally broadened in the second post-compression stage, comprising a reflecting telescope formed by two concave mirrors, each with a 2-m focal length. In the second post-compression stage, another SFC was introduced with entrance and exit diameters of 1 mm and 0.25 mm, respectively. A 0.8-mm-thick fused silica plate was placed 0.74 meters behind the focus point to broaden the injected spectrum to the range of 650-950 nm, where the peak intensity was around 2.3 TW/cm$^2$. The B-integral induced in the second stage was calculated as 3.6. After collimation, the further spectrally broadened laser pulse was dispersion compensated by three pairs of chirped mirrors with a total chirp of -300 fs$^2$. Moreover, the final dispersion compensation also included the dispersion introduced by the subsequent vacuum windows and the customized dichroic mirror used in the synthesizer.

Notably, by optimizing the beam quality injected into the first post-compression stage with the help of a deformable mirror wavefront correction system can effectively improve the overall efficiency of the cascaded compressor, achieving a maximum pulse energy of 20 mJ as the fundamental laser pulses in the synthesizer. Conversely, without the assistance of the wavefront correction system, the output pulse duration of the cascaded compressor remained unchanged, but the pulse energy decreased to 18 mJ.

The temporal profile of the output from the cascaded post-compression was characterized by a home-built SHG-FROG. Given the broad bandwidth of the laser pulse, a BBO crystal with a thickness of merely 10 μm was utilized to generate a sufficiently broadband second harmonic signal. Additionally, an off-axis parabolic mirror was employed to focus the laser pulse under measurement, thereby mitigating the chromatic aberration.

**CEP stability**.

The relative CEP of the Ti:sapphire front-end laser was controlled and locked using a commercial detection module (BIRD, Amplitude) in a feedback loop with an acousto-optic programmable dispersive filter (AOPDF; Dazzler, Fastlite). Fig. 1c presents the measured results of the relative CEP, confirming a stability with a fluctuation of 318 mrad (RMS). Moreover, a single-shot collinear f-to-2f interferometry was applied to monitor the relative CEP value of the injected laser pulse for the HHG experiments. A small portion of laser beam leaking from the back of the dichroic mirror was sent into the f-to-2f interferometry. The fundamental pulse was focused on a 1-mm-thick sapphire plate for spectral broadening across an octave (450-1040 nm). Then, the fundamental pulse was frequency-doubled in the 300-μm-thick β-BBO to generate a second harmonic pulse with the wavelength of 495 nm. After compensation of group delay between the residual fundamental pulse and the auxiliary pulse, both the laser beams passed through the polarizer to generate the interference signal, which was collected by a fiber spectrometer and used to calculate the variation of the relative CEP shift.

**Active stabilization of relative timing delay.**

The relative timing delay between the fundamental (790 nm) and auxiliary (410 nm) laser pulses was actively locked using a home-built balanced optical cross-correlator (BOC). Within the BOC, the fundamental and auxiliary laser pulses leaking from the dichroic mirror were collinearly combined and focused on a 100-μm-thick β-BBO crystal with a cutting angle of 43.6° to generate sum-frequency generation (SFG) signals. By introducing a delay plate, two cross-correlation signals with opposite delay dependencies were produced and subsequently detected by a balanced photo detector. The differential error signals from the balanced photo detector were highly sensitive and strictly proportional to the relative timing jitter between the fundamental and auxiliary laser pulses. This differential error signal was continuously fed into a proportional-integral-derivative (PID) controller, which provided an active feedback voltage to drive the PZT (P-625, Physik Instrumente) stage located in the auxiliary laser path. Consequently, the relative time delay of the two-color synthesizer was tightly locked, reducing the timing jitter to 116 as RMS and guaranteeing the long-term reproducibility of the synthesized waveform.

**All-optical characterization of IAPs**.

The all-optical FROG was employed to characterize the temporal profile of the generated IAPs in Kr and Ar. During the all-optical FROG, the residual fundamental pulse from the frequency-doubling process of the auxiliary laser was utilized as the perturbing laser pulse. A fused silica plate, featuring a broadband anti-reflection coating on one side, was employed to recombine the perturbing pulse and the synthesized driving pulse. The perturbing beam was reflected off the uncoated surface, enabling continuous tuning of its reflectivity by adjusting the incidence angle of the combiner. A PZT (P-625, Physik Instrumente) was introduced in the perturbing beam to precisely scan the relative time delay between the perturbation laser pulse and the two-color synthesizer pulse. In the experiment, to ensure stabilization, a Mach-Zehnder (MZ) interferometer was adopted to stabilize the variations of the optical paths among the perturbing pulse and the two-color synthesizer pulse [42].

The intensity of the perturbation laser pulse was carefully attenuated with an iris to induce a phase modulation on the continuous high harmonic emission without causing excessive ionization of the target medium. The Principal Component General Projection Algorithm (PCGPA) was utilized to retrieve the spectrogram trace of the all-optical FROG. The retrieval errors were 1.1% RMS (Fig. 4b), 1.8% RMS (Fig. 7b) and 1.0% RMS (Fig.8b) in the experiments conducted with Kr, Xe and Ar, respectively.

**Experimental setup for HHG.**

The variable-length gas cell (maximum length of 500 mm) was mounted on a one-dimensional linear translation stage with a 300-mm stroke. This configuration allowed scanning of different parameters related to the phase-matching, including gas pressure, interaction length, and the position of the focal spot. After the interaction between the driving laser and gas medium, a 0.2-μm-thick aluminum film was used to filter out the residual pump pulse. Moreover, a slit located 7 meters away from the HHG position was used to reduce the intensity of the high harmonic pulse injected into the flat-field grating (Shimadzu, featuring 1200 lines/mm) for spectral resolution. Subsequently, the resulting spectrum was measured by a grating-based spectrometer equipped with a microchannel plate (MCP) and a fluorescent screen, which was imaged and recorded by a CCD camera. During the calibration of pulse energy of IAPs, the X-ray CCD was utilized to replace the MCP, enabling direct recording of the photon count.

**Energy characterization of IAPs**.

Based on equation 2, we evaluated the efficacy of each optical element along the detection beam path. Specifically, $\eta_g$ corresponds to the diffraction efficiency of the grating, $T_f$ represents the experimentally determined transmittance of the aluminum (Al) filter, and $T_s$ denotes the transmission efficiency of the slit. Through this meticulous procedure, the count of actual photons corresponding to each pixel on the X-ray CCD can be clearly ascertained. Taking a photon energy of 40 eV in the experiments of Kr as an example, the measured parameters are $S_{CCD}$ = 112062, σ = 2, $\eta_{QE}$ = 68.8%, $\eta_g$ = 3.4%, $T_f$ =15%, and $T_s$ = 2.1%, yielding photon count of $2.78\times10^8$ per shot.

To verify the absolute accuracy of the energy yields measured by the X-ray CCD method, a cross-calibration was performed using an independent method with a photodiode (AXUV100G, Opto Diodes). The cross-calibration was performed for the HHG driven by the fundamental laser pulse. We positioned the photodiode downstream of the grating to exclusively detect the current, which is then converted into an average power with the known responsivity specified in the product datasheet of the photodiode. For instance, by integrating the measured current of the photodiode, we determined the corresponding pulse energy after the grating to be 7.7 picojoules per shot for the 21st harmonic. Under identical experimental conditions, the energy measured by the X-ray CCD method was 8 picojoules. The discrepancy between these two methods was less than 4%. Moreover, the same cross-calibration approach was employed to calibrate the 19th and 23rd harmonics, with errors also remaining below 4% for both.

## Acknowledgments

We acknowledge funding from the National Natural Science Foundation of China (No.12225406) and National Key Research and Development Program of China (No. 2023YFA1406800). We also thank Jingyi Huang for supporting SHG-FROG, and Tongxin Liu for supporting f-to-2f interferometry.

## Author contributions

P.F.L. and P.X.L. conceived this research. L.X. realized the TW few-cycle laser pulse. L.H.M., X.S.Z and L. X. performed the experiments. L.X, L.H.M and P.F.L. analyzed the data and wrote the manuscript. All the authors participated in the discussions and manuscript revise.

## Data availability

All the data that support the findings of this study are available from the corresponding author upon reasonable request.

## Conflict of interest

The authors declare no competing interests.